\documentclass[12pt]{article}
\begin{document}

\begin{center}
{\Large\bf Classical pseudotenors and positivity in small regions}
\end{center}

\begin{center}
$^{1}$Lau Loi So, $^{2}$James M. Nester and $^{3}$Hsin Chen\\
$^{1,3}$Department of Physics, National Central University,
Chung-Li 320, Taiwan.\\
$^{2}$Dept. of Physics and Institute of Astronomy, National Central
University, Chung-Li 320, Taiwan.
\end{center}

\begin{abstract}
We have studied the famous classical pseudotensors in the small
region limit, both inside matter and in vacuum.  A recent work
[Deser $et$ $al.$ 1999 $CQG$ {\bf 16}, 2815] had found one
combination of the Einstein and Landau-Lifshitz expressions which
yields the Bel-Robinson tensor in vacuum.  Using similar methods we
found another independent combination of the Bergmann-Thomson,
Papapetrou and Weinberg pseudotensors with the same desired
property.  Moreover we have constructed an infinite number of
additional new holonomic pseudotensors satisfying this important
positive energy requirement, all seem quite artificial.  On the
other hand we found that M{\o}ller's 1961 tetrad-teleparallel
energy-momentum expression naturally has this Bel-Robinson property.
\end{abstract}

\section{Introduction: quasilocal quantities for small regions}
The localization of gravitational energy-momentum remains an
important problem in GR.  Using standard methods many famous
researchers each found their own expression.  None of these
expressions is covariant, they are all reference frame dependent
(referred to as ``pseudotensors''). This feature can be understood
in terms of the equivalence principle: gravity cannot be detected at
a point, so it cannot have a point-wise defined energy-momentum
density. Now there is another way to address the difficulty.

The new idea is quasilocal: energy-momentum is associated with a
closed 2 surface surrounding a region \cite{Szabados}. A good
quasilocal approach is in terms of the Hamiltonian \cite{cqg}. Then
the Hamiltonian boundary term determines the quasilocal quantities.
In fact this approach includes all the traditional pseudotensors
\cite{prl,gc}. They are each generated by a superpotential which can
serve as special type of Hamiltonian boundary term.

A good energy-momentum expression for gravitating systems should
satisfy a variety of requirements, including giving the standard
values for the total quantities for asymptotically flat space and
reducing to the material energy-momentum in the appropriate limit.
No entirely satisfactory expression has yet been identified.  One of
the most restrictive requirements is positivity. A general
positivity proof is very difficult.  One limit that is not so
difficult is the small region limit.  This has not previously been
systematically investigated for all the classical pseudotensors
expressions.  In matter the expression should be dominated by the
material energy-momentum tensor.  In vacuum we require that its
Taylor series expansion in Riemann normal coordinates have at the
second order a positive multiple of the Bel-Robinson tensor.  This
will guarantee positive energy in small vacuum regions.


\section{The classical pseudotensors}
Recall the Einstein field equation
$G_{\mu\nu}=R_{\mu\nu}-\frac{1}{2}g_{\mu\nu}R=\kappa{}T_{\mu\nu}$.
Matter energy-momentum has a vanishing covariant divergence
$\nabla_{\mu}T^{\mu}{}_{\nu}=0$, but in curved spacetime this is not
in the form of
 a conserved energy-momentum relation. But one can rewrite it in the
 form of a divergence
\begin{equation}
\partial_{\mu}\sqrt{-g}(T^{\mu}{}_{\nu}+t^{\mu}{}_{\nu})=0,
\end{equation}
here $t^{\mu}{}_{\nu}$ is the gravitational energy-momentum
pseudotensor. The energy-momentum complex $\cal{}T^{\mu}{}_{\nu}$ is
then given as
\begin{equation}
\kappa{\cal{}T}^{\mu}{}_{\nu}
=\kappa\sqrt{-g}(T^{\mu}{}_{\nu}+t^{\mu}{}_{\nu})
\equiv\partial_{\lambda}U_{\nu}{}^{[\mu\lambda]},
\end{equation}
where $U_{\nu}{}^{[\mu\lambda]}$ is called the superpotential.

\section{Riemann normal coordinates and the adapted tetrads}

For the total quantities of an isolated gravitating system the
various expressions give the expected weak field asymptotic values.
However they are quite different in the strong field region. To
study the quasilocal quantities one can Taylor expand the
Hamiltonian, including the divergence of its boundary term, in a
small spatial region surrounding a point. The reference is the flat
space geometry at this origin.  Riemann normal coordinates (RNC)
satisfy
\begin{eqnarray}
g_{\mu\nu}(0)=\eta_{\mu\nu},
\quad{}\partial_{\lambda}g_{\mu\nu}(0)=0,\quad
\partial^{2}{}_{\mu\nu}g_{\alpha\beta}(0)
=-\frac{1}{3}(R_{\alpha\mu \beta \nu}+R_{\alpha \nu\beta\mu})(0).
\end{eqnarray}
Later we will also need the associated adapted orthonormal frame
(aka tetrad, vierbein) which satisfies
$g_{\mu\nu}=\eta_{ab}e^a{}_\mu e^b{}_\nu$ and
\begin{equation}
e^{a}{}_{\mu}(0)=\delta^{a}_{\mu}, \quad
\partial_{\nu}e^{a}{}_{\nu}(0)=0,
\quad \Gamma^{a}{}_{b{}\mu}(0)=0, \quad
\partial_{\mu}\Gamma^{a}{}_{b{}\nu}(0)
=\frac{1}{2}R^{a}{}_{b{}\mu\nu}(0).
\end{equation}
For the energy-momentum density we want at non-vacuum points the
results to be dominated by the material energy-momentum.  At vacuum
points one wants a result proportional to the Bel-Robinson tensor.
That will guarantee the proper positive energy property; i.e. the
associated energy-momentum vector will then be future pointing and
non-space like.

\section{Quadratic curvature combinations}
The Bel-Robinson tensor $B_{\alpha\beta\mu\nu}$ and the tensors
$S_{\alpha\beta\mu\nu}$, $K_{\alpha\beta\mu\nu}$ and
$T_{\alpha\beta\mu\nu}$ are defined as follows
\begin{eqnarray}
B_{\alpha\beta\mu\nu}
&:=&R_{\alpha\lambda\mu\sigma}R_{\beta}{}^\lambda{}_{\nu}{}^{\sigma}
+R_{\alpha\lambda\nu\sigma}R_{\beta}{}^{\lambda}{}_{\mu}{}^{\sigma}
+3T_{\alpha\beta\mu\nu}, \\
S_{\alpha\beta\mu\nu}
&:=&R_{\alpha\mu\lambda\sigma}R_{\beta\nu}{}{}^{\lambda\sigma}
+R_{\alpha\nu\lambda\sigma}R_{\beta\mu}{}{}^{\lambda\sigma}
-6T_{\alpha\beta\mu\nu}, \\
K_{\alpha\beta\mu\nu}
&:=&R_{\alpha\lambda\mu\sigma}R_{\beta}{}^{\lambda}{}_{\nu}{}^{\sigma}
+R_{\alpha\lambda\nu\sigma}R_{\beta}{}^{\lambda}{}_{\mu}{}^{\sigma}
+9T_{\alpha\beta\mu\nu},\\
T_{\alpha\beta\mu\nu}&:=&-\frac{1}{24}g_{\alpha\beta}g_{\mu\nu}
R_{\lambda\sigma\xi\kappa}R^{\lambda\sigma\xi\kappa}.
\end{eqnarray}
A recent work \cite{Deser} had found exactly one pseudotensor
expression, a certain combination of the Einstein and
Landau-Lifshitz expressions, which yields the Bel-Robinson tensor in
vacuum.  They argued that this combination is unique under their
assumptions.

\section{Classical holonomic pseudotensors}
The well known classical superpotentials associated with the
Einstein, Landau-Lifshitz, Bergmann-Thomson, Papapetrou, Weinberg
and M{\o}ller(1958) energy-momentum complexes are
\begin{eqnarray}
_{E}U_{\alpha}{}^{[\mu\nu]}
&=&\sqrt{-g}g^{\beta\sigma}\Gamma^{\tau}{}_{\lambda\beta}
\delta_{\tau\sigma\alpha}^{\lambda\nu\mu},\\
_{B}U^{\alpha[\mu\nu]}&=&\sqrt{-g}g^{\alpha\beta}
g^{\pi\sigma}\Gamma^{\tau}{}_{\lambda\pi}
\delta_{\tau\sigma\beta}^{\lambda\nu\mu}=\sqrt{-g}_{L}U^{\alpha[\mu\nu]},\\
_{P}H^{[\mu\nu][\alpha\beta]}&=&
\sqrt{-g}g^{ma}g^{nd}\delta_{ab}^{\mu\nu}\delta_{mn}^{\alpha\beta},\\
_{W}H^{[\mu\alpha][\nu\beta]}&=&\sqrt{-\eta}\left(
\eta^{mc}\eta^{nd}-\frac{1}{2}\eta^{mn}\eta^{cd}\right)
g_{cd}\eta^{ab}\delta_{ma}^{\alpha\mu}\delta_{nb}^{\nu\beta},\\
_{58}U_{\alpha}{}^{[\mu\nu]}
 &=&\sqrt{-g}(\Gamma^{\nu\mu}{}_{\alpha}-\Gamma^{\mu\nu}{}_{\alpha}).
\end{eqnarray}
The pseudotensors are obtained according to one of the
prescriptions:
\begin{equation}
{\cal{}T}_{\alpha}{}^{\mu}=\partial_{\nu}U_{\alpha}{}^{[\mu\nu]},\quad
{\cal{}T}^{\alpha\mu}=\partial_{\nu}U^{\alpha[\mu\nu]},\quad
{\cal{}T}^{\mu\nu}=\partial^{2}{}_{\alpha\beta}H^{[\mu\alpha][\nu\beta]}.
\end{equation}
Inside matter at the origin, the RNC expansion results are
\begin{equation}
E_{\alpha}{}^{\beta}(0)=B_{\alpha}{}^{\beta}(0)=P_{\alpha}{}^{\beta}(0)
=W_{\alpha}{}^{\beta}(0)=2G_{\alpha}{}^{\beta}(0)=2\kappa{}T_{\alpha}{}^{\beta}(0).
\end{equation}
This is as expected from the equivalence principle.  However the
M{\o}ller(1958) expression gives
\begin{equation}
M_{\alpha}{}^{\beta}(0)=R_{\alpha}{}^{\beta}(0)
=\kappa\left(T_{\alpha}{}^{\beta}
-\frac{1}{2}\eta_{\alpha}{}^{\beta}T\right)(0)
\neq{}2\kappa{}T_{\alpha}{}^{\beta}(0).
\end{equation}
This result is not acceptable.  From \cite{Deser}, the small vacuum
($G_{\mu\nu}=0$) region non-vanishing terms  are
\begin{eqnarray}
E_{\alpha}{}^{\beta}&=&-2\Gamma_{\lambda\sigma\alpha}\Gamma^{\beta\lambda\sigma}
+\delta_{\alpha}^{\beta}\Gamma_{\lambda\sigma\tau}\Gamma^{\tau\lambda\sigma},\\
L^{\alpha\beta}&=&\Gamma^{\alpha}{}_{\lambda\sigma}
(\Gamma^{\beta\lambda\sigma}-\Gamma^{\lambda\sigma\beta})
-\Gamma_{\lambda\sigma}{}^{\alpha}(\Gamma^{\beta\lambda\sigma}
+\Gamma^{\sigma\lambda\beta})
+g^{\alpha\beta}\Gamma_{\lambda\sigma\tau}\Gamma^{\sigma\lambda\tau}.\quad
\end{eqnarray}
Using a similar technique we found, for the other pseudotensors
(note Papapetrou and Weinberg depend explicitly on the Minkowski
metric $\eta_{\alpha\beta}$),
\begin{eqnarray}
P^{\alpha\beta}&=&L^{\alpha\beta}
+h^{\lambda\sigma}(\Gamma^{\alpha\beta}{}_{\lambda,\sigma}
+\Gamma^{\beta\alpha}{}_{\lambda,\sigma})
-h^{\lambda\beta}{}_{,\sigma}(\Gamma^{\alpha\sigma}{}_{\lambda}+
\Gamma^{\sigma\alpha}{}_{\lambda}),\\
W^{\alpha\beta}
&=&-2\Gamma_{\lambda\sigma}{}^{\alpha}\Gamma^{\lambda\sigma\beta}
+g^{\alpha\beta}\Gamma_{\lambda\sigma\tau}\Gamma^{\lambda\sigma\tau}
-g^{\alpha\lambda}g^{\beta\pi}
h^{\sigma\rho}\delta_{\lambda\sigma}^{ck}\delta_{\pi\rho}^{\xi{}d}
(\Gamma_{dc\kappa,\xi}+\Gamma_{cd\kappa,\xi}),\quad\quad\\
M_{\alpha}{}^{\beta}&=&\Gamma_{\lambda\sigma\alpha}
\Gamma^{\lambda\sigma\beta}
-\Gamma_{\lambda\sigma\alpha}\Gamma^{\sigma\lambda\beta}+
(g^{\beta\sigma}g^{\psi\lambda}-g^{\beta\psi}g^{\lambda\sigma})
g_{\alpha\psi,\lambda\sigma}.
\end{eqnarray}
Here $h_{\alpha\beta}:=g_{\alpha\beta}-\eta_{\alpha\beta}$ and
$h_{\alpha\beta}=-\frac{1}{3}R_{\alpha\lambda\beta\sigma}x^{\lambda}x^{\sigma}+O(x^{3})$
in RNC.  From \cite{Deser} in vacuum we have
\begin{equation}
\partial^{2}_{\mu\nu}E_{\alpha\beta}(0)
=\frac{1}{9}(4B_{\alpha\beta\mu\nu}-S_{\alpha\beta\mu\nu}),\quad
\partial^{2}_{\mu\nu}L_{\alpha\beta}(0)
=\frac{1}{9}\left(7B_{\alpha\beta\mu\nu}+\frac{1}{2}S_{\alpha\beta\mu\nu}\right).
\end{equation}
In order to obtain the Bel-Robinson tensor, Deser $et$ $al.$ used a
\symbol{92}by hand" combination
\begin{equation}
\partial^{2}_{\mu\nu}\left(\frac{1}{2}E_{\alpha\beta}+L_{\alpha\beta}
\right)=B_{\alpha\beta\mu\nu},
\end{equation}
Using similar methods  we obtained for the other pseudotensors in
vacuum at the origin
\begin{eqnarray}
\partial^{2}_{\mu\nu}P_{\alpha\beta}(0)&=&
\frac{2}{9}(4B_{\alpha\beta\mu\nu}-S_{\alpha\beta\mu\nu}
-K_{\alpha\beta\mu\nu}),\\
\partial^{2}_{\mu\nu}W_{\alpha\beta}(0)&=&
-\frac{2}{9}(B_{\alpha\beta\mu\nu}+2S_{\alpha\beta\mu\nu}
+3K_{\alpha\beta\mu\nu}),\\
\partial^{2}_{\mu\nu}M_{\alpha\beta}(0)
&=&\frac{1}{9}\left(2B_{\alpha\beta\mu\nu}-\frac{1}{2}S_{\alpha\beta\mu\nu}
-K_{\alpha\beta\mu\nu}\right).
\end{eqnarray}
(Here we have included for completeness the result of the
M{\o}ller(1958) expression, even though it does not have a good
matter interior limit.)  From this we find one more independent
combination of the Landau-Lifshitz, Papapetrou and Weinberg
pseudotensors with the same desired Bel-Robinson property. (Note:
the earlier work cited above did not get this result, as they had
excluded the explicit use of the Minkowski metric in the
superpotentials they considered.) Inside matter and in vacuum at the
origin, respectively, we find
\begin{equation}
\frac{1}{3}\left[2L_{\alpha\beta}
+\frac{1}{2}(3P_{\alpha\beta}-W_{\alpha\beta})\right](0)=2G_{\alpha\beta}(0),\quad
\frac{1}{3}\partial^{2}{}_{\mu\nu}\left[2L_{\alpha\beta}
+\frac{1}{2}(3P_{\alpha\beta}-W_{\alpha\beta})\right](0)=B_{\alpha\beta\mu\nu}.
\end{equation}

\section{A large class of new pseudotensors}
Moreover we have constructed an infinite number (a 3 parameter set)
of new holonomic pseudotensors all satisfying this important
Bel-Robinson/positivity property.  The new general superpotential is
\begin{eqnarray}\label{7aDec2005}
{\cal{}U}_{\alpha}{}^{[\mu\nu]}={}
_{E}U_{\alpha}{}^{[\mu\nu]}+\left\{\begin{array}{cc}
c_1h^{\mu\pi}\Gamma_{\alpha}{}^{\nu}{}_{\pi}
+c_2h^{\mu\pi}\Gamma^{\nu}{}_{\alpha\pi}
+c_3h^{\mu\pi}\Gamma_{\pi}{}^{\nu}{}_{\alpha}
\quad\quad\quad\quad\quad\quad\\
+c'_4\delta_{\alpha}^{\mu}h^{\pi\rho}\Gamma^{\nu}{}_{\pi\rho}
+c''_4\delta_{\alpha}^{\mu}h^{\pi\rho}\Gamma_{\pi}{}^{\nu}{}_{\rho}
+c_5h_{\alpha\pi}\Gamma^{\mu\nu\pi} -(\nu\leftrightarrow\mu)
\end{array}\right\},
\end{eqnarray}
where $c_{1}$ to $c_{5}$ are constants and
$h_{\alpha\beta}:=g_{\alpha\beta}-\eta_{\alpha\beta}$. In Riemann
normal coordinates
$h^{\alpha\beta}=\frac{1}{3}R^{\alpha}{}_{\xi}{}^{\beta}{}
_{\kappa}x^{\xi}x^{\kappa}+O(x^{3})$.  Actually, the leading
superpotential $_{E}U_{\alpha}{}^{[\mu\nu]}$ in (\ref{7aDec2005}) is
not necessary Freud's, it can be replaced by any other which offers
a good spatially asymptotic and small region material limit. The
resultant energy density inside matter at the origin is, as
expected,
\begin{equation}
2\kappa{\cal{}E}_{\alpha}{}^{\beta}(0)=2G_{\alpha}{}^{\beta}(0)
=2\kappa{}T_{\alpha}{}^{\beta}(0).
\end{equation}
The RNC second derivatives in vacuum at the origin are
\begin{eqnarray}
\partial^{2}_{\mu\nu}{\cal{}E}_{\alpha\beta}(0)
=\frac{1}{9}\left\{\begin{array}{ccc}
(4-2c_1+c_2+c_3-4c_4+3c_5)B_{\alpha\beta\mu\nu} \\
-\frac{1}{2}(2-c_1-4c_2+5c_3-2c_4-3c_5)S_{\alpha\beta\mu\nu} \\
+(c_1+c_2-2c_3-2c_4)K_{\alpha\beta\mu\nu} \quad\quad\quad\quad\quad
\end{array}\right\}.
\end{eqnarray}
where $c_{4}=c_{4}'-\frac{1}{2}c''_{4}$.
Note that we can choose
\begin{eqnarray}
4-2c_1+c_2+c_3-4c_4+3c_5>0, \\
2-c_1-4c_2+5c_3-2c_4-3c_5=0, \\
c_1+c_2-2c_3-2c_4=0.
\end{eqnarray}
The solutions can be parameterized as follows,
\begin{equation}
c_1+c_2-2c_3<1,\quad
(c_{1},c_2,c_3,c_4,c_5)=\left(c_1,c_2,c_3,\frac{1}{2}(c_1+c_2+2c_3)
,\frac{1}{3}(2-2c_1 -5c_2+7c_3)\right).
\end{equation}
Then the second derivatives of the new pseudotensors in vacuum are
\begin{equation}\
\partial^{2}_{\mu\nu}{\cal{}E}_{\alpha\beta}(0)
=\frac{2}{3}(1-c_{1}-c_{2}+2c_{3})B_{\alpha\beta\mu\nu}.
\end{equation}
As there are three arbitrary constants that one can tune, obviously
there exists an infinite number of solutions with any positive
magnitude of $B_{\alpha\beta\mu\nu}$.
They all appear to be highly artificial.  It seems that there is no
obstruction in going to higher order. From our analysis we infer
that there are an infinite number of holonomic gravitational
energy-momentum pseudotensor expressions which satisfy the highly
desired small region Bel-Robinson/positive energy property.

\section{M{\o}ller's 1961 tetrad-teleparallel energy-momentum tensor}
On the other hand, unlike the aforementioned mathematically and
physically contrived expressions, we found that M{\o}ller's 1961
teleparallel-tetrad energy-momentum expression naturally has the
desired Bel-Robinson property.
The superpotential has the same form as Freud's but the indices now
refer to a tetrad:
\begin{equation}
_{61}U_{a}{}^{[bc]}
=\sqrt{-g}g^{df}\Gamma^{i}{}_{fe}\delta_{ida}^{bce}.
\end{equation}
Expressed in terms of differential forms we have
\begin{equation}
m_{a}{}^{b}\eta_{b} =d(\Gamma^{b}{}_{c}\wedge\eta_{b}{}^{c}{}_{a}).
\end{equation}
The RNC and adapted frame expansion results inside matter, and the
vacuum second derivatives at the origin, respectively, are
\begin{eqnarray}
m_{a}{}^{b}(0)=2G_{a}{}^{b}(0)=2\kappa{}T_{a}{}^{b}(0),\quad
\partial^{2}_{\mu\nu}m_{ab}(0)=\frac{1}{2}B_{ab\mu\nu}.
\end{eqnarray}
Thus the desired Bel-Robinson property is naturally satisfied. An
important consequence is that the gravitational energy according to
this measure is positive, at least to this order. (We expected this
positivity result since in fact M{\o}ller's 1961 expression has an
associated positive energy proof \cite{jmn}.) Once again M{\o}ller's
1961 tensor stands out as one of the best descriptions for
gravitational energy-momentum.


\begin{thebibliography}{3}

\bibitem{Szabados}
L.~B. Szabados, Living Rev. Relativity, {\bf 7} (2004) 4;
http://www.livingreviews.org/lrr-2004-4

\bibitem{cqg}
  C.~M.~Chen and J.M.~Nester,
  {\it Class.{} Quantum Grav.{}}  {\bf 16} (1999) 1279
  [arXiv:gr-qc/9809020].

\bibitem{prl}
  C.~C.~Chang, J.~M.~Nester and C.~M.~Chen,
  {\it Phys.{} Rev.{} Lett.{}} {\bf 83} (1999) 1897.


\bibitem{gc} C.~M.~Chen and J.~M.~Nester,
  {\it Grav.{} \& Cosmol.{}}  {\bf 6} (2000) 257 [arXiv:gr-qc/0001088].




\bibitem{Deser}
S. Deser, J.~S. Franklin and D. Seminara, {\it Class.{} Quantum
Grav.} {\bf 16} (1999) 2815.


\bibitem{jmn}  J.~M.  Nester,
{\it Int.{} J. Mod.{} Phys. A. \bf 4} (1989) 1755.
\end{thebibliography}
\end{document}